\documentstyle[12pt,axodraw]{article}
 
\newcommand{\pr}{Phys. Rev. \underline}
\newcommand{\prl}{Phys. Rev. Lett. \underline}
\newcommand{\pl}{Phys. Lett. \underline}
\newcommand{\np}{Nucl. Phys. \underline}
\newcommand{\ebar}{\overline{\epsilon}}
\newcommand{\msbar}{\overline{\mbox{\tiny MS}}}
\newcommand{\MSbar}{\overline{\mbox{\small MS}}}
\newcommand{\mbar}{\overline{m}} 
\newcommand{\ksi}{{\cal E}} 

\newcommand{\lbar}{\overline{\Lambda}}
\newcommand{\lqcd}{\Lambda_{{\rm QCD}}}
\newcommand{\beq}{\begin{equation}}
\newcommand{\eeq}{\end{equation}}
\newcommand{\be}{\begin{equation}}
\newcommand{\ee}{\end{equation}}
\newcommand{\beqn}{\begin{eqnarray}}
\newcommand{\eea}{\end{eqnarray}}
\newcommand{\bea}{\begin{eqnarray}}
\newcommand{\eeqn}{\end{eqnarray}}

\newcommand{\tnasymp}{t_{n,{\mathrm asymp}}}

\newcommand{\la}{\langle}
\newcommand{\ra}{\rangle}

\newcommand{\sinveff}{S^{-1}_{{\rm eff}}}

\def\spose#1{\hbox to 0pt{#1\hss}}
\def\ltapprox{\mathrel{\spose{\lower 3pt\hbox{$\mathchar"218$}}
 \raise 2.0pt\hbox{$\mathchar"13C$}}}
\def\gtapprox{\mathrel{\spose{\lower 3pt\hbox{$\mathchar"218$}}
 \raise 2.0pt\hbox{$\mathchar"13E$}}}
\def\inapprox{\mathrel{\spose{\lower 3pt\hbox{$\mathchar"218$}}
 \raise 2.0pt\hbox{$\mathchar"232$}}}

\begin{document}
 
\pagestyle{empty}
\begin{flushright}
CERN-TH/96-117\\
ROME prep. 1149/96\\ 
SHEP 96-11 \\
\end{flushright}
 
\centerline{\bf }
\vskip 0.8cm
\centerline{\bf ON THE DIFFICULTY OF COMPUTING HIGHER-TWIST CORRECTIONS}
\vspace{1cm}
\centerline{\bf G.Martinelli$^a$ and C.T.Sachrajda$^{b,*}$}
\vspace{0.4cm}
\centerline{$^a$ Dip. di Fisica, Universit\` a degli Studi di Roma
``La Sapienza'' and }
\centerline{INFN, Sezione di Roma, P.le A.Moro 2, 00185, Rome, Italy}
\centerline{$^b$ Theory Division, CERN, 1211 Geneva 23, Switzerland.}

\vspace{1cm}
\begin{abstract}
  We discuss the evaluation of power corrections to hard scattering
  and decay processes for which an operator product expansion is
  applicable. The Wilson coefficient of the leading-twist operator is
  the difference of two perturbative series, each of which has a
  renormalon ambiguity of the same order as the power corrections
  themselves, but which cancel in the difference.  We stress the
  necessity of calculating this coefficient function to sufficiently
  high orders in perturbation theory so as to make the uncertainty of
  the same order or smaller than the relevant power corrections. We
  investigate in some simple examples whether this can be achieved.
  Our conclusion is that in most of the theoretical calculations which
  include power corrections, the uncertainties are at least comparable
  to the power corrections themselves, and that it will be a very
  difficult task to improve the situation.
\end{abstract}

\vskip 0.5 cm
\begin{flushleft}
CERN-TH/96-117\\
May 1996 \\
\end{flushleft}
\vskip 0.5 cm
$^*$ On leave from the Dept. of Physics,
University of Southampton, Southampton SO17 1BJ, UK.

\newpage
\pagestyle{plain}
\setcounter{page}{1}

\section{Introduction}
\label{sec:intro}

In this paper we address the problem of controlling power corrections
in effective theories. As an example consider $e^+ e^-$ annihilation
into hadrons, for which the cross section is described by a
perturbation series, computed at the parton level, plus power
corrections which are proportional to the condensates of higher
dimensional operators.  In practice the value of the gluon condensate
is obtained by comparing the experimental value of some quantity
derived from $R_{e^+ e^-}(Q^2)$ to its theoretical expression. This
parameter is then used to predict many other physical quantities, such
as form factors and decay constants. Given that only a few (typically
one or two) terms of the perturbative series are known, and that the
series are plagued by renormalon ambiguities, which are of the same
order as the contribution from the condensate, one may wonder whether
the value of the condensate is really known to sufficient accuracy to
be used in other processes where power corrections are important for
the theoretical predictions. This problem is not limited to the gluon
condensate, but is also present for other important parameters of
effective theories, such as the binding energy ($\lbar$) and kinetic
energy ($\lambda_1$) of the Heavy Quark Effective Theory (HQET), and
the matrix elements of higher-twist operators in deep inelastic
scattering (DIS), such as those for the Gross-Llewellyn Smith and
Bjorken sum rules. We argue that the problem is not solved at present,
and that the uncertainties in the determination of these
non-perturbative parameters are seriously underestimated.

The computation of power corrections requires the evaluation of the
matrix elements of higher-twist or higher-dimensional operators.  In
addition, however, it also requires the calculation of the Wilson
coefficient functions to sufficiently high order of perturbation
theory for the cancellation of ``renormalon ambiguities'' to be under
control\,\footnote{We will show in section \ref{sec:power} that, in
  predictions for physical quantities, the Wilson coefficient
  functions can be written as linear combinations of two (or more)
  perturbation series, each of which has a high order behaviour such
  that its Borel transform has a renormalon singularity. The residue
  of the singularity cancels in the combination, however.}. The reason
for this requirement is that these ambiguities are of the same order
as the power corrections.  Since, in the calculations performed up to
now, only the first few terms of the perturbation series are known, it
is not possible to check that the remaining terms are indeed
negligible. By studying some simple examples, we will show that the
knowledge of only a few terms is, in general, insufficient to control
the power corrections. Although the results rely on some
approximations which we are forced to adopt in these examples, it is
likely that our conclusions will remain valid in general, and that in
most cases a further theoretical effort is needed.

Our assumption throughout this paper is that one is attempting to
evaluate the first power corrections (i.e. the next-to-leading twist
contributions\,\footnote{We will frequently misuse the expression
  ``higher-twist'' to mean generic power corrections, and not just
  those to light-cone dominated quantities, such as deep inelastic
  structure functions.}) 
with an uncertainty that is smaller than
these corrections. The discussion can be readily generalized to
include higher-order power corrections, with an increase in technical
complexity, but following the same conceptual principles.

Renormalon singularities, and their implications for theoretical
predictions, have been studied for some time now, and many of the
ingredients of our discussion below can be found in
refs.\cite{background}, and in the work of Mueller~\cite{mueller} in
particular. Recently the effects of renormalons in
predictions for the spectroscopy and decays of heavy quarks, obtained
using the heavy quark expansion, have been
studied~\cite{bsuv}-\cite{ms}.  We summarize in section 2 the picture
which has emerged from these references for the appearance and
cancellation of renormalon ambiguities, without reproducing the
underlying arguments and derivations\,\footnote{ It should be stressed
  that although these arguments, based on the renormalization group,
  analyticity, and/or explicit calculations in the large $N_f$ limit
  ($N_f$ is the number of light quark flavours), are compelling, they
  nevertheless do not constitute a formal proof.}.  Our aim in this
paper is to examine critically the procedure necessary to calculate
the power corrections, and in particular to investigate whether
higher-twist effects are sufficiently under control that their
inclusion reduces the uncertainties in the theoretical predictions.
Our conclusion is that to reach a precision of the order of the power
corrections is likely to be a formidable task, and that the main
limitation comes from the truncation of the perturbation series for
the Wilson coefficients at low orders. The presentation below extends
and clarifies that of our earlier paper~\cite{ms}.

We particularly wish to stress that, although some of the examples
presented below are given using the lattice spacing as the cut-off,
the problems discussed in this paper are completely general, and are
not due to some peculiarity of the lattice regularization. An example,
where the same problems are encountered as in the lattice theory, is
given by the zero recoil inclusive sum rules \cite{bsuv2,klwg}. In
this approach, in order to derive a bound on $\lambda_1$, an
ultraviolet cut-off, $\Delta$, is introduced on each side of the
relation between the time-ordered product of two currents saturated
with hadronic states and the corresponding quantity computed on quark
states using the HQET.  It is not surprising that perturbative
corrections $\sim\,\alpha_s(\Delta)\,\Delta^2$ appear in the bound for
$\lambda_1$, essentially eliminating the predictive power of the
approach \cite{klwg}. The appearance of such power divergences, and
the consequent loss of precision when they are subtracted in a low
order of perturbation theory, are general features in the evaluation
of power corrections \cite{ms}.  A more detailed discussion of this
point will be given in subsec.~\ref{subsec:nonpert}.

The plan of the remainder of the paper is as follows. In the next
section we briefly review the appearance of renormalon ambiguities in
the Wilson coefficient functions of operator product expansions. The
matrix elements of the higher twist or higher dimension operators have
to be evaluated non-perturbatively. This can be done by comparing the
theoretical expression for a physical quantity to its experimental
value (where this is known) or by some non-perturbative method (such
as lattice simulations). We show that in both cases the evaluation of
power corrections requires the calculation of the perturbation series
to sufficiently high orders for the cancellation of renormalon
ambiguities to be under control (sections \ref{subsec:experiment} and
\ref{subsec:nonpert} respectively). In
sections~\ref{sec:toy}-\ref{sec:condensate} we study some simple
examples, in order to investigate numerically the precision that might
be reached in evaluating power corrections. We start by considering a
toy model, which contains many of the general features concerning the
appearance and cancellation of renormalon singularities in operator
product expansions (section~\ref{sec:toy}); we then proceed to the
mass of a heavy quark, or equivalently the binding energy $\lbar$
(section~\ref{sec:pole}), and to the determination of the gluon
condensate and its use in phenomenological applications
(section~\ref{sec:condensate}).  Finally section~\ref{sec:concs}
contains our conclusions.

\section{Power Corrections} 
\label{sec:power}

In this section the cancellation of renormalon ambiguities in the
evaluation of hard scattering and decay processes is
discussed~\cite{background}.

Consider an operator $\hat P$ whose matrix element $\la f|\hat P
|i\ra$ contains the non-perturbative effects for some physical process
${\cal P}$. In general $\hat P$ is non-local, for example it may be
the $T$-product of two electromagnetic or weak currents at small
separations (as in the $e^+ e^-$ annihilation cross-section or weak
decays), or almost light-like separations (as in deep inelastic
structure functions). In these cases $\hat P$ is expanded as a series
of local operators, whose coefficients decrease as powers of the
separation. In some important applications to heavy quark physics, QCD
composite operators (represented by $\hat P$) containing the field of
the heavy quark are expanded in terms of local operators of the HQET,
with coefficients that decrease as inverse powers of the mass of the
heavy quark.

The expansion of $\hat P$ (or the Fourier transform of $\hat P$) in
terms of local operators $O_1(\mu),\, O_2(\mu)$ etc., renormalized at
a scale $\mu$, takes the form:
\be
{\cal P}_{fi}(Q^2) \equiv \la\, f\,|\,\hat P(Q^2)\,|\,i\,\ra = 
C_1(Q^2/\mu^2)\, 
\la\, f\,|\,O_1(\mu) \,|\,i\,\ra
+ \frac{C_2(Q^2/\mu^2)}{Q^n}\,
\la\, f\,|\,O_2(\mu)\,|\,i\,\ra + O\left(\frac{1}{Q^{n+p}}\right)
\label{eq:ope}\ee
where $n\ge 1$ and the coefficient functions $C_i$ are independent of
the states $|i\ra$ and $|f\ra$. $Q$ is a large momentum scale; for
example it may be the centre of mass energy in $e^+e^-$ annihilation,
the momentum transfer in deep inelastic scattering or the
renormalization scale (of the order of the mass of the heavy quark) of
the local operator $\hat P$ in heavy quark matrix elements. For
clarity of notation, throughout this paper we suppress the dependence
of the coefficient functions on the coupling constant, $\alpha_s(Q^2)$.
We assume here that there is only one operator in each of the
first two orders of the expansion. If this is not the case, then there
is an additional mixing of operators, which requires only a minor
modification of the discussion below. We will therefore not consider
this possibility further. The final term of $O(1/Q^{n+p})$ in
eq.(\ref{eq:ope}) represents the contributions of operators of even
higher dimension which will not be discussed here.

In some cases the operator $O_2$ is protected from mixing under
renormalization with $O_1$, because of the presence of some symmetry.
An important example of this in heavy quark physics is the
chromomagnetic operator $\bar h\sigma_{\mu\nu}G^{\mu\nu}h$ (where $h$
is the field of a static quark and $G^{\mu\nu}$ is the gluon field
strength tensor), which cannot mix with the lower dimensional operator
$\bar h h$ because it has a different spin structure. In such cases
the corresponding problem of renormalon singularities does not arise.
For the remainder of this paper we assume that $O_2$ and $O_1$ have
the same quantum numbers, so that they can mix under renormalization.
Another important exception to our general discussion is the
difference of matrix elements of the kinetic energy operator taken
between different hadronic states. In this case the higher dimensional
operator, $\bar h \vec D^2h$, can mix with the lower dimensional one,
$\bar hh$, but as the latter is a conserved current it has the same
matrix element between all single hadron states. Thus the
corresponding renormalon ambiguity cancels in the difference of matrix
elements. In all of these exceptions the matrix elements of the higher
dimensional operators (or linear combinations of matrix elements) are
also the leading contribution to some physical quantity (such as the
$B^*$-$B$ mass splitting in the case of the chromomagnetic operator).

The usefulness of the operator product expansion in eq.(\ref{eq:ope})
comes from the fact that the non-perturbative effects in the process
${\cal P}$ are contained in the matrix elements of the local operators
$O_i$, whereas the coefficient functions $C_i$ are calculable in
perturbation theory. This fundamental property is threatened, however,
by the presence of renormalon singularities in the Borel transform of
the perturbation series of the coefficient functions, as we now
explain.

We start by assuming that the operators on the right-hand side of
eq.(\ref{eq:ope}) are renormalized in some scheme based on the
dimensional regularization of ultraviolet divergences, such as the
$\MSbar$ scheme. Then the perturbation series for the coefficient
function $C_1$ is divergent, and moreover is not Borel-summable, due
to the presence of (infra-red) renormalon singularities in its Borel
transform. Hence the ``sum" of this series is not 
unique, the
ambiguity being of $O(1/Q^n)$. This ambiguity is cancelled by that in
the matrix element of the operator $O_2$, which arises as a result of
an ultraviolet renormalon singularity in the Borel transform of the
perturbation series of its matrix elements. This implies that
renormalization schemes based on dimensional regularization do not
define higher-twist or higher-dimensional operators, such as $O_2$,
unambiguously, and alternative definitions have to be used. We now
consider the two possible alternative approaches in turn. In the first
of these, the problem of the ambiguity in the matrix elements of $O_2$
is eliminated by directly relating two, or more, physical quantities
to which the matrix element of $O_2$ contributes (see section
\ref{subsec:experiment}). In the second, the operator $O_2$ is defined
using a hard ultraviolet cut-off and its matrix element is evaluated
using some non-perturbative method, such as lattice simulations (see
section \ref{subsec:nonpert}).

\subsection{Defining The Matrix Elements of $O_2$ Directly in Terms of 
Physical Quantities}
\label{subsec:experiment}

In this subsection we show how to eliminate the renormalon
ambiguity from the matrix elements of the higher-twist operator $O_2$
by sacrificing the possibility of making a theoretical prediction for
one process, ${\cal P}_{fi}(Q^2)$ say. This is achieved by using 
its experimentally measured value, which we denote by 
${\cal P}_{fi}^{exp}(Q^2)$. It is assumed
here that the matrix element of $O_1$ is already known from some
non-perturbative calculation, or from a second experimental
measurement, or because, as happens in some important cases, it is a
conserved operator or even the identity operator. We now show that
such a procedure requires the control of the difference of two series
which are not Borel-summable up to a precision of $O(1/Q^n)$.  Imagine
that we have computed the series for $C_1$ and $C_2$ in perturbation
theory up to, and including, the term of $O(\alpha_s^k(Q))$. We take $k$
such that the coefficients of the series for $C_1$ are not already
diverging because of the renormalon singularity.  
Then we can define
the matrix element of $O_2$ by the relation
\be
{\cal P}_{fi}^{exp}(Q^2) =
C_1^{(k)}(Q^2/\mu^2)\, 
\la\, f\,|\,O_1(\mu) \,|\,i\,\ra
+ \frac{C_2^{(k)}(Q^2/\mu^2)}{Q^n}\,
\la\, f\,|\,O_2^{(k)}(\mu)\,|\,i\,\ra \ ,
\label{eq:opek}\ee
where the superscript $(k)$ on the coefficients denotes the fact that the
perturbative series for $C_1$ and $C_2$ have been truncated at
$O(\alpha_s^k(Q))$, and in the case of the matrix element of the
higher-twist operator $\la f|O_2^{(k)}(\mu) |i\ra$ that its value,
derived from a physical measurement combined with a perturbative
calculation, depends on the order $k$.  The definition of $\la
f|O_2^{(k)}(\mu) |i\ra$, given in eq.(\ref{eq:opek}) is, of course,
totally unambiguous, although it does depend on $Q$ and on the choice
of process ${\cal P}_{fi}(Q^2)$, as well as on $k$. Different choices
of these parameters can change the value of the matrix element of
$O_2$ by terms of $O(\lqcd ^n)$ (or even larger terms if $k$ is too
small), i.e. by an amount which is of the same order as the expected
value of the matrix element itself.  We assume throughout this paper
that the value of $\alpha_s$ is known to the required accuracy\,\footnote{
Although this may present additional difficulties in practice.}.
All the elements in
eq.(\ref{eq:opek}) are known except for the matrix element of $O_2$:
${\cal P}_{fi}^{exp}(Q^2)$ from experimental measurement, $\la f|
O_1(\mu)|i\ra$ by symmetry, measurement or some non-perturbative
method, and the remaining factors by perturbation theory up to
$O(\alpha_s^k(Q^2))$. We now wish to use the value of 
$\la f|O_2^{(k)}(\mu) |i\ra$
defined in this way to make a prediction for another physical process,
${\cal R}_{fi}(Q^2)$ say\,\footnote{For simplicity we assume here that
  the scale $Q^2$ is the same in both processes; see below for the
  general case.}, up to and including corrections of $O(1/Q^n)$:
\be
{\cal R}_{fi}(Q^2) =
D_1(Q^2/\mu^2)\, 
\la\, f\,|\,O_1(\mu) \,|\,i\,\ra
+ \frac{D_2(Q^2/\mu^2)}{Q^n}\,
\la\, f\,|\,O_2(\mu)\,|\,i\,\ra  + \cdots  \ .
\label{eq:oper}\ee
Returning to the example of $R_{e^+ e^-}(Q^2)$ considered before, this
corresponds to extracting the value of the gluon condensate from the
cross-section at charmonium energies (the process ${{\cal P}}$), 
and using it together with parton model perturbative calculations to 
make predictions for other processes. 
If we know $D_1$ and $D_2$ at order $k$, then
using eq.(\ref{eq:opek}) as the definition for $\la f|O_2^{(k)}(\mu) |i\ra$
we may write
\be
{\cal R}_{fi}(Q^2) =  D_1^{(k)}\la f | O_1 | i \ra 
+ \frac{D_2^{(k)}}{Q^n} \la f|O_2^{(k)} |i\ra + \left[
D_1^{(>k)} - C_1^{(>k)}\frac{D_2^{(k)}}{C_2^{(k)}}\right]
\la f | O_1 | i \ra + O\left(\frac{\alpha_s^{k+1}}{Q^{n}}\right),
\label{eq:rmatel}\ee
where we have suppressed the renormalization scale ($\mu$) and the
arguments of the coefficient functions and of $\alpha_s$. The superscript
$(>k)$ denotes that the perturbation series for the coefficient
function starts from the $O(\alpha_s^{k+1})$ term. Each of the terms
on the r.h.s. of eq.(\ref{eq:rmatel}) is free of renormalon
ambiguities: the coefficients $D_1^{(k)}$ and $D_2^{(k)}$ because they
correspond to series that have been truncated at a finite order in
$\alpha_s$; the matrix element $\la f|O_2^{(k)}(\mu) |i\ra$ because it
has been defined directly from an experimental measurement using
eq.(\ref{eq:opek}); and the coefficient in the third term because the
renormalon ambiguity cancels in the difference of the two series up to
a precision of $O(1/Q^{n+p})$ (the perturbation series for the
coefficient function $D_1$ also has a renormalon ambiguity of
$O(1/Q^n)$, which is cancelled by that in the series for
$C_1(D_2\,/C_2))$. The low order terms in these two series are in
general very different, but the divergent behaviour at high orders is
controlled by the same renormalon singularity and hence is the same.

The traditional procedure is to assume implicitly that the third term
on the r.h.s. of eq.(\ref{eq:rmatel}) is small, and to neglect it.  For
the small values of $k$ for which perturbative results are generally
available, there is in general no guarantee that the third term
cannot give a contribution which is comparable to, or even larger
than, that of the condensate itself. Moreover, the relative size of
the two contributions will depend on the order $k$, the scale $\mu$, 
and the process ${\cal P}_{fi}(Q^2)$ used to define the condensate, as
well as on the process ${\cal R}_{fi}(Q^2)$ for which we want to make
the prediction. 

We have seen that in order for the prediction for ${\cal R}_{fi}(Q^2)$
to be accurate up to and including terms of $O(1/Q^n)$, the
perturbative series for the coefficient functions must be known up to
a sufficiently high order $k$, so as to make the third term on the
r.h.s.  of eq.(\ref{eq:rmatel}) negligible. Formally, each term in
these series is exponentially larger than the power corrections which
are being calculated, and it may take a large number of terms before
the required precision is achieved. In sections
\ref{sec:toy}-\ref{sec:condensate} we investigate the accuracy of the
theoretical prediction as a function of the order in several simple
cases.

In the above discussion we have assumed for simplicity that the two
processes ${\cal P}_{fi}(Q^2)$ and ${\cal R}_{fi}(Q^2)$ occur at the
same large momentum scale $Q^2$ \footnote{We also assume that the
  coefficient functions for the two processes are known at the same
  order $k$.}.  This clearly is not necessary. For example one might
wish to predict the behaviour of the moments of deep inelastic
structure functions with the photon's momentum $Q^2$, using an
experimental measurement at a single value of $Q^2$ to determine the
matrix element of $O_2$. Similar cancellations of renormalon
ambiguities occur also in these cases, because the $Q^2$ dependence of
the coefficient functions is given by perturbation theory.

\subsection{Non-Perturbative Computation of the Matrix Elements of $O_2$}
\label{subsec:nonpert}

In this subsection, the procedure needed to 
compute the matrix elements of the higher-twist operator
$O_2$ (non-perturbatively) is described. This requires the
operators $O_2$ to be defined using a hard (dimensionful) ultraviolet
cut-off $\Lambda$. For example, in lattice simulations it is natural to
use bare operators, defined by the lattice action and with $\Lambda =
a^{-1}$, where $a$ is the lattice spacing. We therefore present the
corresponding discussion in terms of bare operators, so that the 
renormalization scale $\mu$ is replaced by the cut-off $\Lambda$. We
also have in mind that $\Lambda^2 \ll Q^2$, otherwise the use of the
operator product expansion would not be necessary, one could just compute
${\cal P}_{fi}(Q^2)$ directly
\footnote{Although, in some cases this may not be possible in
  Euclidean space.}.
With a hard cut-off the operator $O_2$ mixes with $O_1$, with mixing
coefficients which diverge as $\Lambda^n$. 
Thus, for the process 
${\cal P}_{fi}(Q^2)$, we have
\be
{\cal P}_{fi}(Q^2) =
C_1^{(k)}(Q^2/\Lambda^2)\, 
\la\, f\,|\,O_1(\Lambda) \,|\,i\,\ra
+ \frac{C_2^{(k)}(Q^2/\Lambda^2)}{Q^n}\,
\la\, f\,|\,O_2(\Lambda)\,|\,i\,\ra \ ,
\label{eq:opelk}\ee
where the perturbation series, up to order $k$, for 
the coefficient function $C_1$ takes the form
\be 
C_1^{(k)}(Q^2/\Lambda^2) = c_1^{(k)}(Q^2/\Lambda^2) + 
\widetilde c_1^{\, (k)}(Q^2/\Lambda^2) 
\left(\frac{\Lambda}{Q}\right)^n
\label{eq:c1pert}\ee
and the elements of the series $c_1$ and $\widetilde c_1$ diverge at
most as powers of $\log (Q^2/\Lambda^2)$. The term proportional to
$\widetilde c_1$ arises as a result of the mixing of $O_2$ with $O_1$
\cite{ms}.  The matrix elements of $O_2$, computed with the hard
cut-off, such as the lattice spacing, are well defined and unambiguous.
Thus the same must be true for $C_1$, and the series in
eq.(\ref{eq:c1pert}) is indeed free of ambiguities of $O(1/Q^n)$.
However this arises as a cancellation of the renormalon singularities
in the two series $c_1^{(k)}$ and $\widetilde c_1^{\,(k)}$ as $k$
becomes large.  Each of these two series has a renormalon ambiguity of
$O(1/Q^n)$. The cancellation of this ambiguity occurs between
contributions which, in each order of perturbation theory, are of
different order in $1/Q$. Again the low order terms in the two series
are very different from each other, but the high order behaviour is
governed by the same renormalon singularity and is the same.  In order
to predict ${\cal P}_{fi}(Q^2)$, we have to control the two series in
eq.(\ref{eq:c1pert}) up to a precision of better than
$O(\lqcd^n/Q^n)$.

A related practical problem is the cancellation of the power
divergences of $O(\Lambda^n)$ in the coefficient function $C_1$ with
those in the matrix element $\la\, f\,|\,O_2(\Lambda)\,|\,i\,\ra$.
$C_1$ has to be computed to sufficiently high order so that in
eq.(\ref{eq:opelk}) this cancellation also occurs with a precision of
better than $O(\lqcd^n/Q^n)$.

In addition to the ultraviolet cut-off necessary to regularize the
theory, it is useful in some processes to introduce a physical
cut-off. This is the case, for example, in the zero-recoil sum rules
\cite{bsuv2, klwg}, where the cut-off, $\Delta$, is defined to
suppress the contribution of states with excitation energies greater
than $\Delta$~\footnote{ Note that $\Delta\gg\lqcd$ in order to be
  able to use perturbation theory in the parton sector; on the other
  hand we want $\Delta$ to be as small as possible in order to
  suppress the contribution of excited states and to reduce the
  uncertainty due to the perturbative contributions which are
  quadratic in $\Delta$ (see below).} . There is a parallel between
the above discussion with the hard cut-off $\Lambda$, introduced for
the regularization, and that with the physical cut-off $\Delta$.  In
the latter case, when, for example, the matrix element $\la\,
f\,|\,O_2(\Lambda)\,|\,i\,\ra$ is the kinetic energy $\lambda_1$
(which is of $O(\lqcd ^2)$\,), the corresponding power correction to
the coefficient function (i.e. the term proportional to $\widetilde
c_1^{\,(k)}/Q^2$ in eq.(\ref{eq:c1pert})) contains perturbative
corrections of $O(\alpha_s(\Delta)\,\Delta^2)$. The difficulty in
achieving an accurate determination or bound with a physical cut-off
follows the same discussion as for the hard ultraviolet cut-off.
Only by arriving at an order such that $\alpha_s^n(\Delta)\le
O(\lambda_1/\Delta^2)$, can one obtain significant results.

\subsection{Summary}
\label{subsec:summary}

Before concluding this section, we briefly summarize the main points
of the above discussion.
In order to evaluate the power corrections to hard scattering
processes, it is necessary to determine the matrix elements of the
higher-twist operators, such as $O_2$ in the above examples.  This can
be done by comparing a theoretical prediction which depends on a
matrix element of $O_2$ to the experimenta data, or by computing the
matrix element of $O_2$ using some non-perturbative method. We have
argued that, in the predictions for physical quantities, renormalon
ambiguities, which are of the same order as the power corrections
being evaluated, cancel in the combinations of coefficient functions
given in eqs.(\ref{eq:rmatel}) and (\ref{eq:c1pert}). Either of these
procedures reduces the ``intrinsic'' error of the calculation from
$O(1/Q^n)$ to $O(1/Q^{n+p})$, where by intrinsic we mean the minimum
error achievable in principle. In order to reach the required
precision, however, the series need to be evaluated to a sufficiently
high order. An indication of whether the order is sufficiently high
is given by the ``common-sense'' criterion that the last known term of
the perturbative series of the leading coefficient functions should be
significantly smaller than the power corrections. This point will be
further illustrated in the next section, using a toy example, and in
sections~\ref{sec:pole} and \ref{sec:condensate}, where the presence
and cancellation of renormalon ambiguities in quantities depending on
the mass of the heavy quark and on the gluon condensate will be
discussed.

\section{A Toy Model}
\label{sec:toy}

In this section we study a simple example, which contains many
of the general features expected in operator product expansions,
including the next-to-leading twist contributions, and the 
corresponding renormalon singularity. In this example, the 
``physical'' quantity ${\cal P}$ is defined by 
\be
{\cal P}(x,\ebar)\equiv 10^3 \int_0^\infty du\, e^{-u/x} f(u) \ ,
\label{eq:borel}\ee
where 
\be
f(u) = \frac{1}{1-2u} - \frac{\epsilon^{(1-2u)}}{1-2u}\frac{1}{\Gamma(1+2u)}
\ ;
\label{eq:fu}\ee
$f(u)$ is the Borel transform of ${\cal P}(x, \ebar)$, and $\ebar$ is a
function of $\epsilon$ and $x$ defined in eq.(\ref{eq:ebar}) below.
The factor of $10^3$ in eq.(\ref{eq:borel}) is introduced for
convenience. In eq.(\ref{eq:fu}) the first term, expanded in powers
of $u$ and integrated over $u$ as in eq.(\ref{eq:borel}), generates
the perturbation series of the leading coefficient function in $x$
\beqn
 \int_0^\infty du\, e^{-u/x} \frac{1}{1-2u} & \to &
 \int_0^\infty du\, e^{-u/x} \sum_{j=0}^{\infty} (2u)^j \nonumber\\ 
&  =&
x + 2x^2 +\cdots + \frac{\Gamma(k)}{2} (2x)^k + \cdots \ .
\label{eq:pertser}\eeqn
The second term in eq.(\ref{eq:fu}) corresponds to the matrix element
of the higher-twist operator, and we shall refer to its contribution
to ${\cal P}$ as the ``condensate'' contribution. Its strength is
governed by the parameter $\ebar$, defined by
\be
\epsilon = \ebar\, e^{-1/2x}\ ,
\label{eq:ebar}\ee
which shows that $\epsilon$ is a term of order $\lqcd/Q$ when the
``coupling'' $x\sim 1/\ln(Q^2/\lqcd^2)$~\footnote{Similar models can
  be constructed to mimic the cases in which the power corrections are
  suppressed as $(\lqcd/Q)^n$, with $n>1$.}.  In physical cases,
$\ebar$ is determined by the non-perturbative dynamics, so that ${\cal
  P}$ is only a function of $x$. In our toy model, however, we
will treat $\ebar$ as a free parameter. Its natural value is of
$O(1)$. The factor of $1/\Gamma(1+2u)$ has been introduced in order to
make the integral in eq.(\ref{eq:borel}) converge at large values of
$u$ (for small values of $\ebar$) or to improve the convergence. Other
choices of such damping factors at large $u$ would have been equally
good for our purposes.

Both terms in eq.(\ref{eq:fu}) exhibit a renormalon singularity at
$u=1/2$; this however cancels in $f(u)$, in a way which is analogous
to the cancellation that is expected to occur in physical cases.
Because of this singularity, the coefficients of the perturbation
theory grow like a factorial. As a consequence, even for small values
of $x$, the contribution of high orders diverges. We denote by
$k_{min}$ the order in $x$ for which the magnitude of the $k$-th term
of the series is the smallest one;  $k_{min}$ depends on $x$ only.

In order to mimic the procedure one is forced to adopt in realistic cases,
we proceed as follows:
\vspace{-0.15in} 
\begin{enumerate}
\item[i)] For any choice of $x$ and $\ebar$, the value of ${\cal
    P}(x,\ebar)$ is obtained exactly from eq.(\ref{eq:borel}) by
  numerical integration, and represents the ``experimental'' result
  (with no error).
\item[ii)] We assume that only $k$ terms, with $k \le k_{min}$, of the
perturbation series (\ref{eq:pertser}) are known. 
\item[iii)] We define the ``condensate'' ${\cal C}_k$ as 
\be
{\cal C}_k(x,\ebar) = {\cal P}(x, \ebar) - \frac{1}{2}\sum_{j=1}^k\,
\Gamma(j)\, (2x)^j\ ,
\label{eq:condk}\ee
where by writing explicitly the arguments of ${\cal C}_k$ we recall
that $\ebar$ is being treated as a free parameter. The subscript $k$
is a reminder that the condensate was defined by subtracting $k$ terms
of the perturbation series.
\item[iv)] It is then envisaged that the values of the condensate obtained
in this way are used to make predictions for another process, ${\cal R}$
say, as explained in subsection~\ref{subsec:experiment}.
\end{enumerate}
\begin{table}
\centering
\begin{tabular}{|c|c|c||c|c|c|c|}\hline
\multicolumn{3}{|c||}{\mbox{}}& $\ebar = 100$ & $\ebar = 10$ & 
$\ebar = 1$ & $\ebar = 0.1$ \\ \hline
\multicolumn{3}{|c||}{${\cal P}(x=0.07,\ebar)$}
& 72.23 & 82.33 & 84.92 & 199.57 \\ 
\hline\hline
$k$ & $t_k$ & $s_k$ & \multicolumn{4}{c|}{``Value of Condensate'' -- ${\cal 
C}_k(x,\ebar)$}\\ \hline
1 & 70   & 70    & 2.23    & 12.33 & 14.92 & 129.57 \\ 
2 & 9.8  & 79.8  & $-$ 7.57  & 2.53  & 5.12  & 119.77 \\ 
3 & 2.74 & 82.54 & $-$ 10.32 & $-$0.21 & 2.38  & 117.02 \\ 
4 & 1.15 & 83.70 & $-$ 11.47 & $-$1.36 & 1.23  & 115.87 \\ 
5 & 0.65 & 84.34 & $-$ 12.12 & $-$2.01 & 0.58  & 115.23 \\ 
6 & 0.45 & 84.79 & $-$ 12.57 & $-$2.46 & 0.13  & 114.77 \\ 
7 & 0.38 & 85.17 & $-$ 12.95 & $-$2.84 & $-$0.25 & 114.39 \\ 
8 & 0.37 & 85.55 & $-$ 13.32 & $-$3.21 & $-$0.62 & 114.02 \\ \hline
\end{tabular}
\caption{Perturbative and non-perturbative contributions to ${\cal P}$,
for $x = 0.07$. The full result for ${\cal P}$ is given in the second row.
$t_k$ and $s_k$ are the $k$-th term and the sum of the first $k$ terms
of the perturbation series (\protect\ref{eq:pertser}). The condensate
contributions ${\cal C}_k$ are defined in eq.(\protect\ref{eq:condk}).}
\label{tab:condensate}\end{table}
The stability of the value of ${\cal C}_k$ with $k$, and the
comparison of its value to the $k$-th term of the perturbation series,
monitors the precision that can be reached in predictions for
physical quantities containing corrections of $O(\lqcd/Q)$.  As an
illustration we present in table~\ref{tab:condensate} the numerical
results for $x=0.07$ (for which $k_{min} = 8$) and for $\ebar = 100,
10, 1$ and 0.1\,. The ``physical'' value of ${\cal P}(x=0.07,\ebar)$
for each of the values of
$\ebar$ is given in the second row of the table.  $t_k$ is the $k$-th
term of the perturbation series (\ref{eq:pertser}) and $s_k$ is the
sum of the first $k$ terms, and they are tabulated up to $k=k_{min}=
8$. The remaining entries in the table are the values of the
``condensates'', ${\cal C}_k$, which one would deduce at each order of
perturbation theory. We now comment on these, distinguishing between
large, intermediate and small values of $\ebar$ in turn:
\vspace{-0.15in}
\begin{enumerate}
\item[i)] \underline{Large value of $\ebar$} ($\ebar$ = 100):\\
  For $\ebar$ = 100, the values of the condensate (${\cal C}_k$)
  stabilize quickly, so that already for $k=3$, ${\cal C}_3$ is
  significantly larger than the corresponding term in perturbation
  theory ($t_3$) and the sum of the higher order terms from $t_4$ to
  $t_{k_{min}}$. If this is the case for both ${\cal P}$ and
  ${\cal R}$ (in the notation of section \ref{subsec:experiment})
  then the precision of the prediction for ${\cal R}$ is clearly
  improved by including the contribution from the condensate, provided
  that at least three terms in the perturbation series have been
  calculated. 
\item[ii)] \underline{Intermediate value of $\ebar$} ($\ebar$ =10 and 
$\ebar$ = 1):\\
  As $\ebar$ is decreased, one has to calculate more terms of the
  perturbation theory before the values of ${\cal C}_k$ stabilize.
  For $\ebar = 10$, the improvement in including the condensate
  contribution is, at best, marginal, even if five or six terms of the
  perturbation theory have been computed. For $\ebar = 1$, ${\cal P}$
  is estimated accurately by perturbation theory, and the condensate
  contribution is too small to be determined. In many
  practical situations it may be sufficient to know that the
  condensate is smaller than some value. The uncertainty in the 
perturbation series is independent of the value of $\ebar$.
\item[iii)] \underline{Small value of $\ebar$} ($\ebar = 0.1$):\\
  For small values of $\ebar$ the situation becomes very unstable, and
  the contribution of perturbation theory, even if one includes all
  the terms up to $t_{k_{min}}$, is a poor approximation to ${\cal
    P}$.  The reason can be understood by considering the series in
  $x$ generated by each of the two terms in eq.(\ref{eq:fu}). The
  coefficients of each of these two series grow like the factorial of
  $k$, but this factorial growth is cancelled in their sum. For very
  small values of $\epsilon$, however, this cancellation will only
  begin to take effect at very high orders, in particular at values of
  $k$ such that $k \gg k_{min}$, at which the perturbation series
  (\ref{eq:pertser}) is already diverging rapidly with $k$. It would
  be fascinating to find a realistic physical example corresponding to
  this case.
\end{enumerate}

The conclusions which we draw from this simple example are as follows.
Imagine that ${\cal P}(x, \ebar)$ has been measured, and that $k$ terms of
the perturbation series have been calculated. A reasonable estimate of
the perturbative contribution to ${\cal P}$ would then be $s_k \pm
t_k$. If ${\cal C}_k \gg t_k$, then it makes sense to call $C_k$ the
condensate contribution, and to use it in predictions for other
processes for which the perturbation series has similar properties and
has been calculated to the same precision. The challenge in
phenomenological applications is to demonstrate that this is the case.
Otherwise the condensate is of the same order as the uncertainty in
the perturbation series (or smaller) and hence its value would depend
on the process from which it is extracted.  $t_k$ serves as an
estimate of the uncertainty in the value of the condensate.

There is one further important point which we wish to stress, i.e. the
universality of the higher order corrections. Consider a set of
processes ${\cal P,R}\cdots$ for which the leading power correction is
given by the same matrix element of a given higher-dimensional operator.
The high order behaviour of all the leading coefficient functions is
then dominated by the same infra-red renormalon, and is hence
universal.  This leads to the possibility that the error in the
prediction for process ${\cal R}$, obtained by using the condensate
${\cal C}_k$ determined in process ${\cal P}$, may be smaller than the
estimated uncertainty in the value of the condensate ${\cal C}_k$ itself.
In order to see this, note that
the definition of the condensate in eq.(\ref{eq:condk}) implies that
\be
t_{k+1}^{\cal P} + {\cal C}_{k+1} - {\cal C}_k = 0\ ,
\label{eq:tkplus1}\ee
where the superscript on $t_{k+1}^{\cal P}$ implies that this is the
$(k+1)$-th term in the perturbation series for the process ${\cal P}$.
The difference between the $(k+1)$-th order and $k$-th order
predictions for ${\cal R}$ is 
\be
t_{k+1}^{\cal R} + {\cal C}_{k+1} - {\cal C}_k + O\left(x \frac{\lqcd}
{Q}\right)\ ,
\label{eq:rpred}\ee
where the last term represents perturbative, renormalon-free
corrections to the power suppressed term. If the perturbation series
for the two processes are both dominated by the same renormalon, and
$k$ is sufficiently large, then it may be that $t_{k+1}^{\cal R}\simeq
t_{k+1}^{\cal P}$, and that the $(k+1)$-th contribution to the
prediction for ${\cal R}$ is smaller than the difference in the
condensates, ${\cal C}_{k+1} - {\cal C}_k$. In practice, however, often
only one or two terms of the perturbation series are known and it is
unclear to what extent this property of universality will be useful in
phenomenological applications.

\section{The Pole Mass of a Heavy Quark} 
\label{sec:pole}

In this section the discussion of section \ref{sec:power} is applied
to the computation of the heavy-quark mass in the HQET. Of course in
practice we do not know the perturbation series for the coefficient
functions to sufficiently high order to be able to study the numerical
effects of the cancellation of renormalon ambiguities directly. For
this reason we present a calculation performed in the limit of a large
number of light quark flavours ($N_f$), or more precisely we perform
the perturbative calculations keeping only the term with the highest
power of $\beta_0$ at each order: $\beta_0 = 11 -2/3 N_f$ is the
lowest order coefficient in the $\beta$-function.

It is possible to compute the renormalized mass of a heavy quark,
defined at a large renormalization scale, $\mu\gg\lqcd$, from the 
matrix elements of the HQET obtained with some non-perturbative
method, such as lattice simulations \cite{cgms}. As an example we
consider $\mbar$, the mass defined in the $\MSbar$ scheme at a 
renormalization scale $\mu = \mbar$:
\be
\mbar = m^{\msbar}(m^{\msbar})\ .
\label{eq:mbardef}\ee
The computation requires the expansion of the propagator of the
heavy quark in QCD (with the mass, wave-function renormalization and
coupling constant defined in the $\MSbar$ scheme say), in terms of
matrix elements of operators in the HQET. We will see that the
coefficient function of the leading operator, $C_1$ of
eq.(\ref{eq:ope}), is indeed a difference of two series (as in
eq.(\ref{eq:c1pert})), each of which has a renormalon ambiguity of
$O(\lqcd)$, the ambiguity cancelling in the difference.  In the large
$N_f$ limit the terms of the series can be calculated to arbitrarily
high orders, and the cancellation of the ambiguity then observed. This is
done in subsection \ref{subsec:pv}; in subsection \ref{subsec:lattice}
we start with a brief review of how one computes $\mbar$ using
non-perturbative methods such as lattice simulations.

\subsection{Evaluation of $\mbar$ in Lattice Simulations}
\label{subsec:lattice}

In this subsection the procedure needed to evaluate
$\mbar$ from simulations in the HQET up to, and including, terms of
$O(\lqcd)$, but neglecting terms of $O(\lqcd^2/\bar m)$ is briefly
reviewed \cite{cgms}
\footnote{In ref.\cite{cgms} the generalization of this discussion to
  include terms of $O(\lqcd^2/\mbar)$ is also presented.}.  The
non-perturbative quantity which is computed directly in lattice
simulations, and which is required for the determination of $\mbar$,
is the bare binding energy $\ksi_H$, where the label $H$ denotes the 
hadron containing the heavy quark. $\ksi_H$ is obtained from the time
dependence of the correlation function of two interpolating operators
($J_H$) for the hadron $H$:
\be
\sum_{\vec x}\, \langle 0\,|J_H(\vec x, t)\,J^\dagger_H(\vec 0, 0)\,
|\,0\rangle = Z\, e^{-\ksi_H\,t}\ ,
\label{eq:cf}\ee
where $t$ is sufficiently large for the correlation function to be 
dominated by the lightest particle created by $J_H^\dagger$, which
is assumed to be $H$.

The relation between $\ksi_H$ and $\mbar$ can be obtained by matching
the heavy quark propagator in QCD with operator matrix elements
evaluated in the HQET \cite{cgms}. Using the $\MSbar$ renormalization 
scheme at a scale $\mu$ for the mass, wave function and coupling constant
renormalization, the inverse propagator in QCD ($S^{-1}$) is of the form:
\beqn
\lefteqn{S_P^{-1}(v\cdot k) = m_Q - m(\mu )\sum_{n=0}^{\infty}
\left( \frac{\alpha_s(\mu)}{4 \pi}\right)^n c_n(m(\mu)/\mu)}
\nonumber\\ 
& & + v\cdot k \sum_{n=0}^{\infty}
\left( \frac{\alpha_s(\mu)}{4 \pi}\right)^n d_n(v\cdot k/\mu, m(\mu)/\mu)
+O(\lqcd^2/\mbar)
\label{eq:spm1qcd}\eeqn
where the momentum of the heavy quark is $m_Q v + k$
\footnote{The precise definition of $m_Q$ here can be conveniently
chosen later.};
$v$ is the four velocity of the heavy quark and 
$S_P$ is defined by 
\be
\frac{1+\rlap/ v}{2} S_P  = \frac{1+\rlap/ v}{2} \, S \,
\frac{1+\rlap/ v}{2} \ .
\label{eq:spdef}\ee
The first series
on the right-hand side of eq.(\ref{eq:spm1qcd}) is just the perturbative
expansion of the pole mass of the heavy quark in terms of $m(\mu)$:
\be
m_{{\rm pole}} = m(\mu )\sum_{n=0}^{\infty}
\left( \frac{\alpha_s(\mu)}{4 \pi}\right)^n c_n(m(\mu)/\mu)\ .
\label{eq:mpole}\ee
The Borel transform of the series in eq.(\ref{eq:mpole}) has
renormalon singularities, which is a ma\-ni\-festation of the fact that
the pole mass is not a physical quantity \cite{bsuv,bb}. The ambiguity
corresponding to the leading singularity is of $O(\lqcd)$.

Now consider perturbation theory in the HQET (defined by the action
$ \bar h\, v\!\cdot\! D\, h$), using the lattice spacing
as the ultraviolet cut-off. $S_P^{-1}$ can be expressed in terms of 
$\sinveff$, the propagator in the HQET, which can be interpreted as 
the matrix element of the operator $O_2 = \bar h\, v\!\cdot\! D\, h$:
\beqn
S_P^{-1}(v\cdot k) 
& = & m_Q - \left(m_{{\rm pole}} - \delta m \right)\nonumber \\ 
& + & C_{v\cdot D}(m(\mu)a, m(\mu)/\mu)\sinveff(v\cdot k\, a) 
+O(\lqcd^2/m(\mu ))\ ,
\label{eq:spm1qcd2}\eeqn
where the series 
\be
\delta m = \frac{4}{a}\frac{C_F}{\beta_0}\sum_{n=0}^{\infty}
\left( \frac{\alpha_s(\mu)}{4 \pi}\right)^{n+1} \frac{X_n}{a}
\label{eq:deltam}\ee
is just the perturbative expansion of the inverse of the quark
propagator in the HQET at zero momentum. The leading renormalon
singularity in $m_{{\rm pole}}$ is cancelled by the one in the series for
$\delta m$, so that the combination $m_{{\rm pole}} - \delta m$ has no
renormalon ambiguity of $O(\lqcd)$~\cite{ms}. The two series in 
eqs.(\ref{eq:mpole}) and (\ref{eq:deltam}) are an example of the 
series in eq.(\ref{eq:c1pert}).
Here $\Lambda = a^{-1}$, $Q$ is the mass of the heavy quark and $n=1$,
i.e. we are calculating the $O(\lqcd)$ correction to the mass of the 
heavy quark, which is one power of $m$ smaller
than the leading term. The operator $O_1$ is $\bar hh$, which is a
conserved current in the HQET with matrix element equal to 1, and so 
it does not appear explicitly in eq.(\ref{eq:spm1qcd2}).

The mass $\mbar$ is obtained from the relation
\be
m_{{\rm pole}} - \delta m = M_H - \ksi_H \ ,
\label{eq:mbardetermination}\ee
where $M_H$ is the physical mass of the hadron $H$, and by inverting
the relation in eq.(\ref{eq:mpole}) between the pole and $\MSbar$
masses. In eq.(\ref{eq:mbardetermination}), the linear divergence
present in $\ksi_H$ is cancelled by that in the series $\delta m$, and
the renormalon in $m_{{\rm pole}}$ is cancelled by that in $\delta m$,
as explained above. In the following subsection we study the numerical
cancellation of the renormalon ambiguity between $m_{{\rm pole}}$ and
$\delta m$. Numerical results for $\mbar$ obtained in this way
(but with the perturbative terms only computed to one-loop order) have
been presented in \cite{cgms,gms}.

\subsection{Cancellation of Renormalon Ambiguities in the Heavy-Quark Mass}
\label{subsec:pv}

In this subsection we explicitly trace how the cancellation of the
renormalon ambiguities occurs in the combination $m_{{\rm pole}} -
\delta m$. The calculation is performed in the large-$\beta_0$ limit,
in which only the terms containing the leading power of $\beta_0$ are
kept in each order of perturbation theory. As a further
simplification, we perform this calculation in the large-$\beta_0$
limit, with the Pauli-Villars (PV) cut-off $\Lambda$ as the
ultraviolet regulator in the effective theory (rather than the
lattice spacing)~\footnote{ The general features of the cancellation
  of the renormalon ambiguities are qualitatively the same with any
  hard cut-off. For the purposes of illustration the PV cut-off is
  very convenient.}. In other words, in analogy with lattice field
theory, we imagine that we have computed $\ksi_H$ non-perturbatively
in the PV theory for some hadron $H$, and now perform the calculation
of the matching term $m_{{\rm pole}} - \delta m$ in perturbation
theory.

\begin{figure}[t]
\begin{center}
\begin{picture}(294,40)(-140,-10)
\Line(-140,0)(154,0)
\Line(-140,-2)(154,-2)
\Photon(-100,0)(-70,25){2}{6}
\Oval(-62,25)(8,8)(0)
\Photon(-54,25)(-34,25){2}{4}
\Oval(-26,25)(8,8)(0)
\Photon(-18,25)(2,25){2}{4}
\Text(12,25)[]{$\cdots$}
\Photon(22,25)(42,25){2}{4}
\Oval(50,25)(8,8)(0)
\Photon(58,25)(78,25){2}{4}
\Oval(86,25)(8,8)(0)
\Photon(94,25)(124,0){-2}{6}
\end{picture}
\caption{Diagrams which must be evaluated in order to study the heavy 
  quark mass (and hence $\lbar$) in the large-$\beta_0$ limit. The
  double line represents the propagator of the heavy quark, and the
  bubbles represent light-quark loops.}
\label{fig:deltam}
\end{center}
\end{figure}
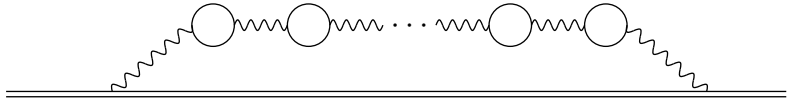

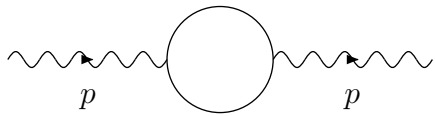
\begin{figure}[t]
\begin{center}
\begin{picture}(160,45)(-80,-25)
\Photon(-80,0)(-20,0){3}{5}\Oval(0,0)(20,20)(0)
\Photon(20,0)(80,0){3}{5}
\ArrowLine(-52,0)(-48,0)\Text(-50,-15)[]{$p$}
\ArrowLine(48,0)(52,0)\Text(50,-15)[]{$p$}
\end{picture}
\caption{One-loop bubble graph contributing to the vacuum polarization of the
  gluon. This graph is the basic ingredient in the evaluation of the
  large-$\beta_0$ contribution to the quark mass (and other
  quantities). The solid lines represent light-quark propagators.}
\label{fig:1loop}
\end{center}
\end{figure}

In order to obtain the result in the lowest non-trivial order in the
large-$\beta_0$ approximation, it is sufficient to evaluate the set of
diagrams in fig.~\ref{fig:deltam}, summing the contributions 
from an arbitrary
number of light-quark loops.  Consider a single light-quark loop
insertion as in fig.~\ref{fig:1loop}. We denote the expression from
this diagram by $L(p, m_q^2)$, where $m_q$ is the mass of the light
quark. $L$ is given in terms of a divergent integral, which we
regulate by taking $L(p,0) + L(p, 2\Lambda^2) - 2 L(p,\Lambda^2)$ as
the regulated expression for the diagram, as suggested in
ref.\cite{iz}.  Other choices are also possible.  For the gluon
propagator it is sufficient to take $-i g^{\mu\nu} (1/q^2 - 1/(q^2 -
\Lambda^2))$ since any terms proportional to $q^\mu q^\nu$ in the
gluon propagator do not contribute to $\delta m$. We then find
\be 
\delta m = 4\Lambda \frac{C_F}{\beta_0}
\sum_{n=0}^{\infty}
\left(\frac{\alpha_s(\Lambda)\beta_0}{4\pi}\right)^{(n+1)} K_n \ ,
\label{eq:deltamresult}\ee
where
\be
K_n = \int^{\infty}_0\frac{dx}{(1 + x^2)^{n+1}}\,
[5/3 - \ln (2x^2) - F(x^2/2) + 2 F(x^2)]^n \ ,
\label{eq:kn}\ee
$\alpha_s(\Lambda)$ is the bare coupling constant and the function 
\be
F(x^2) = 6 \int _0^1 dy \, y(1-y)\, \ln (1 + y(1-y)x^2) 
\label{eq:fdef}\ee
can be readily evaluated. The position of the leading
infra-red renormalon singularity is known, and by inverting the Borel
transform using the saddle point method we find that the behaviour of
$K_n$ at large $n$ is given by: 
\be K_n \to \frac{e^{5/6}}{\sqrt{2}}
2^n n!
\label{eq:knlimit}\ee
The numerical results obtained by using eq.(\ref{eq:kn}) approximate
their asymptotic values in (\ref{eq:knlimit}) to better than 5\%
already for $n=4$.

The relation between the pole mass and $\mbar$ in the large-$\beta_0$
limit is given by \cite{bbb}:
\be
m_{{\rm pole}} = \mbar\left[ 1 + \frac{\alpha_s(\mbar)C_F}{\pi}
\left( 1 + \sum_{n=1}^{\infty} d_n \left(
\frac{\alpha_s(\mbar)\beta_0}{4\pi}\right)^n\right)\right]\ ,
\label{eq:mpoleseries}\ee
where $\alpha_s(\mbar)$ is the $\MSbar$ coupling constant at the 
renormalization scale $\mbar$.
For large values of $n$, the coefficients $d_n$ behave as
\be 
d_n \to e^{5/6} 2^n n!
\label{eq:dnlimit}\ee
and again this asymptotic relation is well satisfied by the numerical results
for relatively small values of $n$.

To demonstrate the numerical cancellation of the leading renormalon
singularity, we do the following. In the second column of table
\ref{tab:masses} we present the results for the pole mass obtained
from eq.(\ref{eq:mpoleseries}), with the assumption that $\mbar = 4.5$
GeV, and $\alpha_s(\mbar)= 0.2$. The results are presented in
successive orders of perturbation theory in the $\MSbar$ coupling
constant at $\mu = \mbar$.  As expected, the value of $m_{{\rm pole}}$
increases rapidly at high orders due to the presence of the renormalon
singularity. The standard approach when dealing with an asymptotic
series is to consider only the first few terms to estimate the result.
Since the smallest term in the series is of $O(100\,{\rm MeV})$ (for
$n=3-5$), i.e.  of $O(\lqcd)$ as expected, this can be viewed as the
intrinsic uncertainty in the unphysical quantity $m_{{\rm pole}}$. In
the third column of table \ref{tab:masses} we present the analogous
results for $m_{{\rm pole}} - \delta m$, also expanded in terms of the
$\MSbar$ coupling constant at $\mu = \mbar$. In
eq.(\ref{eq:deltamresult}) we have taken $\Lambda = 2$ GeV, and have
expanded the Pauli-Villars coupling $\alpha_s(\Lambda)$ in terms of
the $\MSbar$ coupling $\alpha_s(\mbar)$.  The series of the difference
$m_{{\rm pole}} - \delta m$ stabilizes at lower orders because of the
cancellation of the leading renormalon singularity.  The remaining
uncertainty is now much smaller (in general it would be of
$O(\lqcd^2/\mbar)$, but in the large-$\beta_0$ limit the corresponding
renormalon is absent \cite{bb,mns}, so that the ambiguity is of
$O(\lqcd^3/\mbar^2))$.  As a consequence the smallest term in the
series, which sets the scale of the intrinsic uncertainty, is now less
than 10 MeV.

\begin{table}
\centering
\begin{tabular}{|c|c|c|}\hline
$n$ & $m_{{\rm pole}}$ GeV& $m_{{\rm pole}} - \delta m$ GeV \\ \hline
tree & 4.5  &  4.5 \\  
0 & 4.88  &  4.62  \\
1 & 5.12  &  4.72  \\ 
2 & 5.24  &  4.74  \\ 
3 & 5.34  &  4.76  \\
4 & 5.44  &  4.77  \\
5 & 5.58  &  4.77  \\
6 & 5.80  &  4.78  \\
7 & 6.21  &  4.78  \\ 
8 & 7.08  &  4.79  \\ \hline
\end{tabular}
\caption{$m_{{\rm pole}}$ and $m_{{\rm pole}} - \delta m$ calculated
up to $O(\alpha_s^{n+1}(\mbar))$
in perturbation theory, in the large-$\beta_0$ limit, using $\mbar = 4.5$
GeV and $\alpha_s(\mbar) = 0.2$.}
\label{tab:masses}
\end{table}

The unphysical parameter $\lbar=M_H-m_{{\rm pole}}$ is frequently used
in phenomenological studies of $B$-physics. Results and bounds for
$\lbar$ are presented (see for example \cite{lbarresults} and the
reviews \cite{mncp,shifmantasi}, and references therein). In order for
this to make any sense, $\lbar$ must be defined precisely in terms of
some physical quantity, and its value will depend on this physical
quantity and on the order of perturbation theory used to extract
$\lbar$. As can be seen in table~ \ref{tab:masses}, the values for
$\lbar$ will change by several hundred MeV as the order of
perturbation theory is increased.  This also implies that when using
the value of $\lbar$ measured in one process to make predictions for a
second one, we should use the same order of perturbation theory in
both processes and hope that the universality discussed at the end of
section~\ref{sec:toy} holds to a good approximation.

All calculations of power corrections to hard scattering and decay
processes will involve a cancellation of renormalon ambiguities
similar to the one discussed in this section\,\footnote{ Unless there
  is a symmetry which prevents the mixing of the operators with
  different dimensions.}.  Of course, one should remember that the
results in table \ref{tab:masses} were obtained using an approximation
and can only be taken as being indicative. Nevertheless they highlight
the difficulty of evaluating power corrections.  If one's aim is to
try to evaluate $\mbar$ up to uncertainties of $O(\lqcd^2/\mbar)
\simeq 25$~MeV or so, many orders of perturbation theory would be
required because the series for $m_{{\rm pole}} - \delta m$ converges
very slowly. In the lattice theory only the one-loop term is known
(corresponding to $n = 0$ in table \ref{tab:masses}), and the
continuum relation (\ref{eq:mpole}) is known up to two-loop order.
This would suggest that values for the binding energy
$\overline{\Lambda}$ which are used in phenomenological studies in
heavy-quark physics are uncertain by an amount of order 100~MeV, due to
our ignorance of the higher order perturbative terms. A similar comment
applies to our computation of $\mbar$ in ref.\cite{cgms} where this 
uncertainty was probably underestimated, and to the result in \cite{gms}.

It is likely that the case of $\lbar$ is a relatively good one, since
the corrections are suppressed by just one power of the mass of the
heavy quark, and, at least in the large-$\beta_0$ approximation, 
the asymptotic behaviour (\ref{eq:knlimit}) and
(\ref{eq:dnlimit}) seems to set in at low orders of perturbation theory. 
In the next section we consider an important example where this is not the
case, that of the evaluation of the gluon condensate.

\section{The Gluon Condensate}
\label{sec:condensate}

In this section we study another important example, that of the
contribution of the gluon condensate to physical quantities in
general, and to the $D$-function in $e^+e^-$ annihilation in particular.
The $D$-function is defined by $D(Q^2)=-1/4 Q^2 d\Pi(Q^2)/dQ^2$, where
$\Pi$ is obtained from the correlation function of two electromagnetic
currents:
\be
i\int\,d^4x\,e^{iq\cdot x}\la\,0 |\,T\{J_\mu(x)J_\nu(0)\}\,|0\,\ra
= (q_\mu q_\nu - g_{\mu\nu}q^2) \Pi(-q^2)
\label{eq:pidef}\ee
and $Q^2=-q^2$. The gluon condensate, $\langle\,\alpha_s
G^2/\pi\,\rangle$, is the vacuum expectation value of an operator of
dimension 4, and hence its contribution is of $O(\lqcd^4/Q^4)$
relative to the perturbative terms. As we have tried to stress
throughout this paper, the gluon condensate itself is not a physical
quantity, as it contains a renormalon ambiguity of $O(\lqcd^4)$.  This
ambiguity is cancelled by that in the perturbation series for the
$D$-function (or, in general, by that in the leading coefficient
function for the process being studied). The suppression by four
powers of $Q$ implies that the cancellations are very large and leads
to enormous difficulties in the quantitative evaluation of the power
corrections.

In subsection~\ref{subsec:dcont} below we study the high order
behaviour of the perturbative series for the $D$-function in the large
$\beta_0$ limit.  We argue that even at low values of $Q^2$, where the
relative contribution of the power corrections is significant, the
uncertainty and ambiguity in the perturbation series can be comparable
to the contribution normally ascribed to the condensate.  We start,
however, by a discussion of the computation of the gluon condensate in
lattice simulations. We demonstrate that the extremely large numerical
cancellations which arise in the subtraction of the
quartic power divergence (i.e. of the terms which diverge as
$a^{-4}$), and the presence of the corresponding renormalon singularity,
make the quantitative evaluation of the leading power corrections
(i.e. the $O(1/Q^4)$ corrections) prohibitatively difficult.

\subsection{Evaluation of the Gluon Condensate in Lattice Simulations}
\label{subsec:latcond}

A natural definition of the condensate in lattice QCD is given in
terms of the expectation value of the plaquette variable $P_{\mu\nu}$
($\mu$ and $\nu$ define the plane containing the plaquette, but the
expectation value is, of course, independent of the choice of plane):
\be
{\it P} \equiv \la\ 1 - \frac{1}{3} {\mathrm Tr} P_{\mu\nu}\ \ra
\equiv \frac{\pi^2}{36} a^4\, \la\ \frac{\alpha_s}{\pi} G^2 \ \ra _{{\mathrm
latt}}\ ,
\label{eq:latconddef}\ee
where $a$ is the lattice spacing and the subscript stands for
``lattice''. The variable ${\mathit P}$ is measured very
precisely in lattice simulations for the Wilson action, at all
standard values of the lattice spacing (as well as for some other
lattice discretizations of QCD). In lattice QCD, as with any
regularization using a hard cut-off, ${\mathit P}$ is not zero in
perturbation theory, but is given by an expansion of the form
\be
{\mathit P} = \sum_{n=1} \frac{c_n}{\beta^n} \ ,
\label{eq:ppert}\ee
where $\beta = 6/g_0^2(a)$ and $g_0(a)$ is the bare lattice coupling
constant. The series in eq.(\ref{eq:ppert}) arises as a result of the
mixing of the $G^2$ with the identity operator.  The first 8 (!)
coefficients $c_i$ have been obtained numerically using Langevin
techniques \cite{langevin}. We would like to use the computed value of
${\mathit P}$ and perturbative matching to calculate the
$O(\lqcd^4/Q^4)$ corrections to some physical process. For example for
the $D$-function the relation is:
\be
D(Q^2) = D_{\mathrm pert}^{\mathrm cont} - D_{\mathrm
  pert}^{\mathrm latt} + \frac{24}{a^4Q^4} \left(1 + \frac{7}{6}\,\frac
{\alpha_s^{\msbar}(Q)}{\pi}\, + \cdots\right) {\mathit P}\ , 
\label{eq:dexp}\ee
where
\be
D_{\mathrm pert}^{\mathrm cont}  =  
1 + \frac{\alpha_s^{\msbar}(Q)}{\pi} +
1.640 \left(\frac{\alpha_s^{\msbar}(Q)}{\pi}\right)^2 + \cdots
\label{eq:dpertcont}\ee 
and
\be
D_{\mathrm pert}^{\mathrm latt}  = 
\frac{24}{a^4Q^4} \left(1 + \frac{7}{6}\,\frac
{\alpha_s^{\msbar}(Q)}{\pi}\, + \cdots\right)\sum_{n=1} \frac{c_n}{\beta^n}
\ . 
\label{eq:dpertlatt}\ee
$D_{\mathrm pert}^{\mathrm cont}$ is the perturbative series for the
$D$-function and the superscript stands for ``continuum''. The series
$D_{\mathrm pert}^{\mathrm latt}$ arises from the matching of the
$D$-function with the lattice operator in eq.(\ref{eq:latconddef}),
and the superscript stands for ``lattice''.  In the notation of
subsection~\ref{subsec:nonpert}, and eqs. (\ref{eq:opelk}) and
(\ref{eq:c1pert}) in particular,
\be
D_{\mathrm pert}^{\mathrm cont} =  c_1(\Lambda^2/Q^2)\, ,
\label{eq:equivalence1}\ee
and
\be
D_{\mathrm pert}^{\mathrm latt} =  -
\widetilde c_1(Q^2/\Lambda^2)\left(\frac{\Lambda}{Q}\right)^n\, ,
\label{eq:equivalence2}\ee
with $\Lambda = a^{-1}$ and $n=4$. In this case the operator $O_1$ is
the identity operator.  The leading infra-red renormalon singularity
in the series $D_{\mathrm pert}^{\mathrm cont}$ is cancelled by
the renormalon in $D_{\mathrm pert}^{\mathrm latt}$, as explained
in subsection \ref{subsec:nonpert}.

As an example of how serious the cancellations are, and how very
difficult it is to obtain results with the required precision we take
$Q=m_\tau$ ($m_\tau$ is the mass of the $\tau$-lepton), $\beta = 5.7$,
which corresponds to an inverse lattice spacing of about 1.15~GeV, and
$\alpha_s^{\msbar}(m_\tau)$ = 0.32. We are forced to choose fairly large
values of $Q^2$, for which the contribution of the condensate is expected
to be small, since in order to make use of the lattice results we require
$\lqcd<a^{-1}<Q$. We consider $\beta = 5.7$ ($a^{-1}\simeq 1.15\,$GeV) to be
about the smallest value of $\beta$ (and inverse lattice spacing) for which
one may reasonably expect that lattice artefacts will not invalidate
the interpretation of the results. For these values of the parameters,
the three components in eq.(\ref{eq:dexp}) begin like:
\beqn
D_{\mathrm pert}^{\mathrm cont} & = & 
1 + 0.102 + 0.017 + \cdots
\label{eq:dcontnum}\\ 
D_{\mathrm pert}^{\mathrm latt} & = &
5.630 - 6.259 + \cdots
\label{eq:dlattnum}\\ 
\frac{24}{a^4Q^4} \left(1 + \frac{7}{6}\,\frac
{\alpha_s^{\msbar}(Q)}{\pi}\, + \cdots\right) {\mathit P}
& = & (1 + 0.119 + \cdots )\, 1.894 \ .
\label{eq:pnum}\eeqn 
The numbers in eq.~(\ref{eq:dlattnum}) have been obtained after
rewriting the series in terms of $\alpha_s^{\msbar}(Q)$. We see that
the terms in eqs.(\ref{eq:dlattnum}) and (\ref{eq:pnum}) are huge
compared to the contribution one would normally ascribe to the gluon
condensate of about 1\% or so (taking $\langle\,\alpha_s
G^2/\pi\,\rangle \simeq 0.018\,$GeV$^4$). It would clearly be
enormously difficult to quantify these power corrections accurately.
This would require the perturbation series $D_{\mathrm pert}^{\mathrm
  cont} - D_{\mathrm pert}^{\mathrm latt}$, and also the relation
between the lattice and $\MSbar$ coupling constants, to be known to
extremely high orders.

Although the numerical results presented above were obtained using the
plaquette variable to define the gluon condensate on the lattice as in
eq.(\ref{eq:latconddef}), the general discussion applies to any
choice of operator. The problems arise from the presence of quartic
power divergences and renormalon ambiguities, which are general features
of lattice attempts to evaluate the power corrections associated 
with the gluon condensate.

In ref.\cite{ji} the author has attempted to define the gluon
condensate from ${\mathit P} - \sum c_n/\beta^n$, where ${\mathit P}$
has been measured numerically, and the first 8 coefficients $c_i$ are
known\,\cite{langevin}. He uses different resummation techniques for
the perturbative terms and finds results for the condensate which
depend significantly on the method of summation, and are always at
least five times larger than those used in phenomenological
applications. The point that we are trying to stress in this paper is
that such an analysis is theoretically inconsistent.  The series $\sum
c_n/\beta^n$ has a renormalon ambiguity of $O(\lqcd^4)$ (which, as
always, is of the same order as the effect one is trying to evaluate);
thus subtracting the perturbative series $\sum c_n/\beta^n$ from the
computed values of ${\mathit P}$ leaves an intrinsic arbitrariness of
this order. In order to eliminate this arbitrariness we have to
include the measured value of ${\mathit P}$ in the prediction for a
physical quantity using the matching procedure described above. In
this way, for the $D$-function (which is a typical example) the
renormalon ambiguity cancels between the first two terms on the
right-hand side of eq.~(\ref{eq:dexp}).

\subsection{The Uncertainty in the Perturbation Series for the D-function}
\label{subsec:dcont}

Although the discussion in the previous subsection was concerned
specifically with the evaluation of the power corrections to the $D$-function
using lattice
simulations, we believe that it is also very difficult to control the
corresponding calculations in phenomenological studies using continuum
regularizations. Consider the perturbation series for the $D$-function
using the $\MSbar$ coupling constant: \be D_{\mathrm pert}^{\mathrm
  cont} = 1 + \frac{\alpha_s(Q^2)}{\pi} +
1.6398\left(\frac{\alpha_s(Q^2)}{\pi}\right)^2 + 6.37101
\left(\frac{\alpha_s(Q^2)}{\pi}\right)^3 + \cdots \ .
\label{eq:dpert}\ee
The coefficient of $(\alpha_s(Q^2)/\pi)^4$ has been estimated to be 
about 27.5 \cite{k4est}, based on calculations using the principle of minimal
sensitivity \cite{pms} and the effective charge approach \cite{effch}.
For values of $Q^2$ such that $\alpha_s(Q^2) < 1/2$ say, the series appears
to be reasonably well behaved.
However, in order to gain some insight into the effects of the infra-red
renormalon in the series $D_{\mathrm pert}^{\mathrm cont}$ we
have to go beyond the order for which the coefficients are known. For
this reason we study this series in the large-$\beta_0$ limit, for which
the coefficients have been determined in refs.~\cite{broadhurst,beneke}.
Writing
\be
D_{\mathrm pert}^{\mathrm cont}
= 1 + \frac{1}{\beta_0}\sum_{n=1}^\infty\,\kappa_n\,\left(
\frac{\beta_0\alpha_s(Q^2)}{\pi}\right)^n\ ,
\label{eq:kappadef}\ee
the coefficients $\kappa_n$ can readily be obtained from the Borel
transform of the series,
\be
\sum_{n=1}^{\infty} \frac{(4u)^{n-1}}{\Gamma(n)}\, \kappa_n
= \frac{32\, e^{-Cu}}{3(2-u)}\sum_{k=2}^\infty\,\frac{(-1)^k\, k}
{[k^2-(1-u)^2]^2} \ ,
\label{eq:kappagen}\ee
where in the $\MSbar$ scheme $C=-5/3$.  The large order behaviour of
the series $D_{\mathrm pert}^{\mathrm cont}$ is dominated by the
singularity closest to the origin, which in this case is a double
(ultraviolet) renormalon pole at $u=-1$. The high order terms
generated by this pole diverge like a factorial of the order, but with
alternating signs, corresponding to a behaviour which is 
Borel-summable.  Thus although the presence of ultraviolet renormalons may
add further practical difficulties to the evaluation of the power
corrections of $O((\lqcd/Q)^4)$, we will not consider them further
here. Specifically, we subtract the contributions of the ultraviolet
poles at $u=-1$ and $u=-2$, which appear in the terms with $k=2$ and
$3$ respectively in eq.~(\ref{eq:kappagen}), by considering the
behaviour of the coefficients $\kappa^\prime_n$ obtained from
\be
\sum_{n=1}^{\infty} \frac{(4u)^{n-1}}{\Gamma(n)} \kappa^\prime_n
= \frac{32e^{-Cu}}{3(2-u)}\left\{\frac{90-39u+5u^2}{144(3-u)^2} -
\frac{224 - 72u + 7u^2}{576(4-u)^2} + 
\sum_{k=4}^\infty\,\frac{(-1)^k\, k}
{[k^2-(1-u)^2]^2} \right\} \ .
\label{eq:kappapgen}\ee
The residues of the infra-red renormalons (at $u=2$ and above)
are the same in eqs.(\ref{eq:kappagen}) and
(\ref{eq:kappapgen}). The large order
behaviour of the coefficients $\kappa_n^\prime$ is given by
\be
\kappa_n^\prime \rightarrow \kappa_{n,{\mathrm asymp}}^\prime
=\frac{e^{10/3}}{8^{n-1}}\, (n-1)!\ ,
\label{eq:kappaasymp}\ee
as $n\rightarrow\infty$.

Consider the perturbation series generated by the coefficients 
$\kappa_n^\prime$, $D^{\mathrm cont}_{\mathrm pert} = \sum_{n=0} \, t_n$,
where $t_0=1$ and 
\be
t_n = \frac{\kappa^\prime_n}{\beta_0}
\left(\frac{\beta_0\alpha_s(Q^2)}{\pi}\right)^n
\label{eq:tndef}\ee
for $n\ge 1$. We denote by $s_n$ the sum of the series up to $n$-th
order, $s_n=\sum_{k=0}^n t_k $. It is also convenient to define
$t_{n,{\mathrm asymp}}$ as in eq.(\ref{eq:tndef}), but
with $\kappa_n^\prime$ replaced by the asymptotic form $\kappa^\prime
_{n,{\mathrm, asymp}}$.

\begin{table}
\centering
\begin{tabular}{|c|c|c|c|}\hline
$n$ & $s_n$ & $t_n$ &$t_{n,{\mathrm asymp}}$ \\ \hline
0 &  1 & 1 & 1 \\ 
1 & 1.049 &  0.049 & 4.461\\
2 & 1.089 &  0.040 & 0.799\\ 
3 & 1.125 &  0.036 & 0.286\\
4 & 1.161 &  0.036 & 0.154\\
5 & 1.201 &  0.039 & 0.110\\ 
6 & 1.248 &  0.047 & 0.099\\
7 & 1.310 &  0.062 & 0.106\\
8 & 1.399 &  0.090 & 0.133\\ \hline
15 & 23.511 & 13.016 & 13.536\\ \hline
\end{tabular}
\label{tab:coeffs}
\caption{Values of $s_n,\, t_n$ and $t_{n,{\mathrm asymp}}$ for a value
of $Q^2$ such that $\alpha_s(Q^2)=1/2$ ($Q\simeq 0.75$~GeV).}
\end{table}

In order to illustrate the difficulties of evaluating the perturbation
series with sufficient precision to make the inclusion of the
corrections of order $(\lqcd/Q)^4$ meaningful, we present an example.
In table \ref{tab:coeffs} we give the values of $s_n$, $t_n$ and
$t_{n,{\mathrm asymp}}$ obtained for a value of $Q^2$ such that the
$\MSbar$ coupling constant $\alpha_s(Q^2)=1/2$ ($Q\simeq 0.75$~GeV).
We choose a small value of $Q^2$ so as to enhance the contribution of
the power corrections (it would be even more difficult to quantify
the power corrections at higher values of $Q^2$). 
We now make some comments on these results:
\begin{enumerate}
\item[i)] The smallest term in the series $\{ t_n\}$ occurs for $n=4$,
  $t_4=0.036$. One might therefore be tempted to take the four-loop
  result, $s_4 = 1.161$, as the best estimate for the sum of the
  perturbation theory, and $t_4\sim 4\%$ as the estimate of the
  renormalon ambiguity. One might also expect that the inclusion of
  the gluon condensate will eliminate this uncertainty (up to a
  precision of order $(\lqcd/Q)^6$\,). Moreover since with standard
  phenomenological values of the condensate ($\langle\,\alpha_s
  G^2/\pi\,\rangle \simeq 0.018\,$GeV$^4$) it is expected that its
  contribution to the $D$-function at such low values of $Q^2$ is
  about 40\%, the ambiguity of about $4\%$ can be considered
  negligible.
  
  Such an interpretation is wrong, however. The perturbation theory for
  $n\simeq 4$ is not yet dominated by the leading infra-red renormalon
  (as can be seen, for example, from the fact that $t_4$ is very
  different from $t_{4,{\mathrm asymp}}$, or by evaluating the
  contribution from the next-to leading renormalon at $u=3$). The
  smallest term in the series $\{\tnasymp\}$ occurs at $n=6$ and is
  about 10\%. Even for $n\simeq 6$, however, the perturbation series
  $\{t_n\}$ is not well approximated by $\{\tnasymp\}$. Thus it is not
  easy to estimate the uncertainty in the evaluation of the
  perturbation series, other than to say that it is certainly greater
  than 10\%.
  
\item[ii)] Since the perturbation series approaches its asymptotic
  value very slowly, there is no reason why the low order terms should
  be approximately universal. For example, the large order
  contribution to $\kappa^\prime_n$ from the infra-red renormalon at
  $u=3$ is $-4/27\,e^5\,n!/12^{n-1}$ and for $n\le 4$ is greater than
  or comparable to that of the leading infra-red renormalon at $u=2$
  (which is $e^{10/3}\, (n-1)!/8^{n-1}$, see
  eq.(\ref{eq:kappaasymp})\,). The renormalon at $u=3$ corresponds to
  operators of dimension 6, whose contribution relative to the gluon
  condensate depends on the process, and hence the low order terms of
  the perturbation series are not universal\,\footnote{ A related
    question is whether at such low scales, the non-perturbative
    contibutions of the operators corresponding to $u=3$ and above do
    not become as large as that of the gluon condensate.}.
\item[iii)] The discussion in i) and ii) was based on the assumption
  that it is possible to calculate many orders of perturbation theory
  and to study the extent to which the asymptotic behaviour has been
  reached.  Of course, in practice, usually only one or two terms of
  the perturbation theory are known, which adds substantially to the
  uncertainty. For example the values of the contribution from the
  ``condensate'', which one would obtain by subtracting either the
  one-loop result ($s_1$) or the six-loop one ($s_6$) from the measured
  value of $D(Q^2)$, would differ by about 0.2, i.e. by 20\% of the
  $D$-function itself.
\item[iv)] It may be the case that for some processes, and at small
  values of $Q^2$ in particular, the ``condensate'' contribution is
  much larger than all the combined uncertainties. In the language of
  subsection~\ref{subsec:experiment} this would be necessary both for
  the physical quantity ${\cal P}$ being used to determine the
  condensate (e.g. some correlation function used in the study of the
  spectrum of charmonium) and for the quantity ${\cal R}$ for which
  the prediction is being made (e.g. decay constants or semileptonic
  form factors of heavy mesons). This example demonstrates, however,
  that to be confident that the uncertainties are indeed sufficiently
  small will require considerable effort.
\end{enumerate}

In this simple example the uncertainty and ambiguity in the
perturbation series for $D(Q^2)$ is at least a significant fraction 
($50$--$100 \%$ ?) of
the expected size of the leading power correction. This is in spite of
the fact that $Q^2$ was chosen to be small in an attempt to minimize
the relative size of the ambiguity. Although one can change the
details of the discussion by using different values of $Q^2$,
different renormalized coupling constants as the expansion parameters,
or different physical processes, it is our contention that the
difficulties discussed above are general and cannot be easily overcome.

\section{Conclusions}
\label{sec:concs}
In this paper we have studied several examples in order to understand
whether it is possible to compute power corrections to hard scattering
and decay processes to a sufficient level of precision. By this we
mean that the theoretical uncertainties must be smaller than the power
corrections themselves. In all of the examples considered, we have
found that this is not possible unless we are able to control
perturbation theory at higher orders than those available at present.
Since the arguments discussed in this paper are general, and not
specific to the examples used, we believe that even the leading power
corrections are currently not well determined, and there is little
hope to compute higher order power corrections\,\footnote{For
  exceptions to the general discussion see section~\ref{sec:power}.}.
We hope that these disappointing conclusions and provocative comments
will help to stimulate further debate and a systematic investigation
of this central question of particle physics phenomenology.

\section*{Acknowledgements}
We thank J.Ellis, H.Leutwyler, G.Marchesini, P.Nason, M.Neubert,
M.Shifman, M.Testa, N.Uraltsev and A.Vainshtein for many stimulating
discussions and for their encouragement. G.M. thanks G.Veneziano and
members of the Theory Division at CERN for their hospitality during
the completion of this work. G.M. acknowledges partial support from
M.U.R.S.T. and C.T.S. acknowledges the Particle Physics and Astronomy
Research Council for its support through the award of a Senior
Fellowship. We also acknowledge partial support by the EC contract
CHRX-CT92-0051.


\begin{thebibliography}{99}    
\bibitem {background}
G. 't Hooft, in: {\it The Whys of Subnuclear Physics},
ed. A.~Zichichi (Plenum Press, New York, 1979), p.~943;
B. Lautrup, \pl {B69} (1977) 109;
G. Parisi, \pl {B76}  (1978) 65 and
\np {B150} (1979) 163;
F. David, \np {B234} (1984) 237 and
{\it ibid.\/} \underline {B263} (1986) 637;
V.I. Zakharov, \np {B385}, 452 (1992);
M. Beneke and V.I. Zakharov, \prl {69} (1992) 2472;
M. Beneke, \pl {B307} (1993) 154 and \np
{B405} (1993) 424;
D. Broadhurst, Z.\ Phys.\ \underline{C58} (1993) 339;
X.Ji, \np {B 448} (1995) 51
\bibitem{mueller} A.H. Mueller, \np {B250} (1985) 327, \pl {B308}
  (1993) 355 and in the proceedings of the Workshop ``QCD: 20 years
  Later'', Aachen, June 1992, eds. P.M.Zerwas and H.A.Kastrup (World
  Scientific, Singapore, 1993), p.~162;
\bibitem{bsuv} I.I.Bigi, M.A.Shifman, N.G.Uraltsev and A.I.Vainshtein,
\pr {D 50} (1994) 2234
\bibitem{bb} M.Beneke and V.M.Braun, \np {B 426} (1994) 301
\bibitem{bbz} M.Beneke, V.M.Braun and V.I.Zakharov, \prl {73} (1994) 3058 
\bibitem{ns} M.Neubert and C.T.Sachrajda, \np {B 438} (1995) 235
\bibitem{lms} M.Luke, A.V.Manohar and M.J.Savage, \pr {D 51} (1995) 4924
\bibitem{ms} G.Martinelli and C.T.Sachrajda, \pl {B 354} (1995) 423
\bibitem{bsuv2} I.I.Bigi, M.A.Shifman, N.G.Uraltsev and A.I.Vainshtein,
\pr {D 52} (1995) 196
\bibitem{klwg} A.Kapustin, Z.Ligieti, M.B.Wise and B.Grinstein,
Caltech preprint CALT-68-2041 (1996) - (hep-ph/9602262)
\bibitem{cgms} M.Crisafulli, V.Gimenez, G.Martinelli and C.T.Sachrajda,
\np {B 457} (1995) 594
\bibitem{gms} V.Gimenez, G.Martinelli and C.T.Sachrajda,
University of Rome Preprint, in preparation
\bibitem{iz} C.Itzykson and J.-B.Zuber, {\it Quantum Field Theory}, 
McGraw-Hill, New York, 1980
\bibitem{bbb} P.Ball, M.Beneke and V.M.Braun, \np {B 452} (1995) 563
\bibitem{mns} G.Martinelli, M.Neubert, and C.T.Sachrajda,
\np {B 461} (1996) 238 
\bibitem{lbarresults} M.Neubert, \pr {D 46} (1992) 1097;
E.Bagan, P.Ball, V.M.Braun and H.G.Dosch, \pl {B 278} (1992) 457;
A.F.Falk, M.Luke and M.J.Savage, Toronto Preprint UTPT-95-24 (1995)
(hep-ph/9511454);
M.Gremm, A.Kapustin, Z.Ligieti and M.B.Wise, Caltech Preprint CALT-68-2043
(1996) (hep-ph/9603314)
\bibitem{mncp} M.Neubert, CERN Preprint CERN-TH/96-55 (1996) (hep-ph/9604412)
\bibitem{shifmantasi} M.Shifman, Minneapolis Preprint TPI-MINN-95/31-T(1995)
(hep-ph/9510377) 
\bibitem{langevin} F.Di Renzo, E.Onofri and G.Marchesini, 
\np {B 457} (1995) 202
\bibitem{ji} X-D. Ji, MIT preprint MIT-CTP-2439 (1995) (hep-ph/9506413)
\bibitem{k4est} A.L.Kataev and V.V.Starshenko, Mod. Phys. Lett.
  \underline{A 10} (1995) 235
\bibitem{pms} P.M.Stephenson, \pr {D 23} (1981) 2916
\bibitem{effch} G.Grunberg, \pl {B 221} (1980) 70; \pr {D 29} (1984)
  2315
\bibitem{broadhurst} D.Broadhurst, Z. Phys.  \underline{C 58} (1993)
  339
\bibitem{beneke} M.Beneke, \pl {B 307} (1993) 154; \np {B 405} (1993)
  424
\end{thebibliography}
\end{document}